\documentstyle[prd,aps,preprint,epsfig,floats]{revtex}
\tightenlines

\newcommand{\m}{m_e}
\newcommand{\En}{E_{\nu}}
\def\DESepsf(#1 width #2){\epsfxsize=#2 \epsfbox{#1}}
\begin{document}

\draft

%@twocolumnfalse\endcsname
\preprint{\vbox{
\hbox{UMD-PP-04-041}
}}

\title{{\Large\bf Reactor Searches for Neutrino Magnetic Moment as a Probe
of Extra Dimensions}}
\author{\bf  R. N. Mohapatra, Siew-Phang Ng and Haibo Yu}

\address{ Department of Physics, University of Maryland, College Park,
MD 20742, USA}
\date{April, 2004}
\maketitle

\begin{abstract}
We present calculations of the magnetic moment contribution to
neutrino electron scattering in large extra dimension brane-bulk
models (LED) with three bulk neutrinos. We discuss the cases with
two and three large extra dimensions of sizes $R$. The calculations are
done using reactor flux from
Uranium, $^{235}U$ as the neutrino source. We find that if the
electron neutrino mass is chosen to be in the range of one eV, the
differential cross section for $\bar{\nu}-e$ scattering for low electron
recoil energy can be of the same order as the
presently explored values in reactor experiments. Furthermore the
spectral shape for the LED models is different from the four
dimensional case. Future higher precision reactor
experiments can therefore be used to provide new constraints on a
class of large extra dimension theories.
\end{abstract}

\vskip1.0in
\newpage

\section{Introduction}
The possibility that there may be extra hidden small-sized space
dimensions has been the subject of a great deal of attention in
recent years, mostly motivated by string theories as well as
phenomenological considerations having to do with understanding
the gauge hierarchy problem. A specially interesting subclass of
these theories are the ones where the extra space dimensions have
sizes ranging from several micrometers to a millimeter\cite{add}.
They have the attractive phenomenological property that
 they can be tested in laboratory experiments searching
for deviations from inverse square law of gravity. In this note we discuss
one way to test these models using neutrino experiments.

One of the ways to explain the observed neutrino masses and
mixings in large extra dimension (LED) models is to postulate the
existence of a singlet neutrino in the bulk\cite{addm} and couple
it to the known neutrinos on the brane. Since such a particle from
a four dimensional point of view is equivalent to a tower of
Kaluza-Klein(KK) states, it tends to manifest itself in many
experiments even though it has no conventional weak interaction.
One such dramatic manifestation is in the effective magnetic
moment interaction in $\bar{\nu} e$ and $\bar{\nu} N$ elastic
scattering\cite{ng}. The reason for this is that the magnetic
moment couples the left handed neutrino to the full tower of
states. This means that in the magnetic part of $\bar{\nu} e$ scattering
one
can produce all the states in the KK tower
$m_{\nu_R, KK}\leq \En$. All these  final states will add
incoherently to the magnetic moment part of the cross section and
will in general lead to considerable enhancement. One can
therefore use reactor neutrino experiments searching for magnetic
moment\cite{before,future} to constrain the size of the extra
dimensions in the LED models. 

In this brief note, we present a
calculation of the magnetic moment contribution to $\bar{\nu}-e$
scattering by reactor neutrinos. We calculate the flux averaged
$\frac{d\sigma}{dT}$ for a $^{235}U$ neutrino source where $T$ is the
electron recoil energy. We find that for the case of two large
extra dimensions, the $\frac{d\sigma}{dT}$ can be comparable to the
corresponding one for the case where there are no extra dimensions
and the $\mu_\nu\sim 10^{-10}\mu_B$, which was calculated in
\cite{vogel}.The value for $\mu_\nu$ chosen is the present upper
bound from the previous experiments\cite{before}. 
 For the case of three extra dimensions, the value is 
lower.  We further
point out that for the LED models, the spectral shape of
$d\sigma/dT$ is different from the case with a single right
handed neutrino with an appropriate magnetic moment. This can therefore
be a distinguishing signature of the LED models.

\section{Neutrino magnetic moment in $\delta$ large extra dimensions}
For the case of $D$ large extra dimensions and $\delta$ small
dimensions, the size and the fundamental scale are related by the formula:
\begin{eqnarray}
       M^2_{P}~=~M^{2+D+\delta}_{*D} (\pi R)^{D} (\pi r)^{\delta}
\end{eqnarray}
where we have assumed all large dimensions to have the same size $R$ and
all small ones to have the same size $r$. If we take the $M_{*D}$ to be a
TeV, we find that for $D=2$ and $M_{*D}\pi r=1$, $R\sim 2$ mm and for
$D=3$ and
$M_{*D}\pi r=1$, $R=0.02$ $\mu$m. Below, we will present our calculations
for both these cases.

Let us now briefly discuss the minimal neutrino mass scenario for
these models where one introduces three singlet neutrinos in the
bulk\cite{addm,others}. The standard model particles in this
picture reside on the brane whereas the singlet neutrinos $N$ are in
the bulk. One introduces a brane bulk coupling in the form:
\begin{eqnarray}
{\cal L}_N~=~\frac{h_{\alpha\beta}}{(2\pi R)^{D/2}} \int d^Dy
\delta^D(y) \bar{\psi}_{L,\alpha}H
N_{\beta} + h.c.
 \end{eqnarray}
where $R$ is the size of each of the extra space dimensions on which the
theory is compactified. The extra dimensions are compactified on an
$\left(\frac{S_1}{Z_2}\right)^D$ space. We can expand the bulk
neutrinos $N_\beta$ into their Fourier modes as $N_\beta~=~\Sigma
N^{n}_{\beta}(x)\frac{e^{\frac{in.y}{R}}}{\sqrt{(2\pi R)^D}}$  to get
 the neutrino mass of the form
\begin{eqnarray}
{\cal M}_{\alpha \beta}~=~ \Sigma _n h_{\alpha\beta}\frac{v_{wk}
M_{*D}}{M_P}\bar{\nu}_{L,\alpha}N^{(n)}_{R,\beta}
\end{eqnarray}
This matrix can be diagonalized by a common $3\times 3$ rotation
for all the KK modes for the bulk modes i.e.
$N^n_\beta=V^n_{\beta j} N^n_j$ and a rotation on the lefthanded
brane neutrinos $\nu_\alpha~=~U_{i\alpha}\nu_i$. The latter leads
to the PMNS matrix measured in neutrino oscillation experiments.
In this diagonal basis, the mixing of the brane neutrinos to the
$n$th KK mode of the bulk singlets are given by a common factor
$\xi_i~=~\sqrt{2} m_i R/n$. The neutrinos in this model are Dirac
neutrinos.

Let us now discuss the magnetic moment predictions of this model. As
already noted, the magnetic moment will connect the brane neutrino to all
the KK excitations of the bulk singlets equally. For example for the one
neutrino case, the magnetic moment interaction will be given by\cite{fs}:
\begin{eqnarray}
{\cal L}_{\mu_{\nu}}~=~\mu_l\mu_B 
\Sigma_n\bar{\nu}_L\sigma^{\mu\nu}N^{(n)}F_{\mu\nu} 
\end{eqnarray}
where $\mu_l~=~\frac{3G_F m_e
m_\nu}{4\sqrt{2}\pi^2}~=~ 3.2\times
10^{-19}\left[\frac{m_\nu}{1~eV}\right]$; $\mu_B$ is the Bohr
magneton. The symbol $n$ goes over a square or cubic lattice depending 
on whether $D=2$ or $D=3$. Because of this, when we calculate $\bar{\nu} 
e$ scattering using this formula, the
phase space as well as parameters such as the size $R$ will be very
different for the two cases leading to different final results.

\section{Magnetic moment contribution to $\bar{\nu} e$ scattering cross
section}
In this section we calculate the differential cross section as a
function of the electron recoil energy $T\equiv E'-m_e$ ($E'$ is
the final state electron energy). The techniques used are
standard. The only point that need careful treatment is the phase
space and implementation of the energy momentum conservation. For
instance, a naive estimate of the $\frac{d\sigma}{dT}$ for the
case one extra D is to take the four dimensional result and
multiply it by $(\En R)$, where $E_\nu$ is the incident neutrino
energy. That this is not correct can be seen by noting that the
above estimate assumes that all KK states up to $E_\nu$
contribute; however energy momentum conservation allows only modes
$\ll \En $ to contribute. Furthermore, energy momentum conservation
imposes other constraints.

The calculated cross-section for $\bar{\nu_e}+e\rightarrow e+\bar{N}$
where N is a right handed neutrino of mass M is,
\begin{equation} \frac{d \sigma}{d T}= \frac{\pi \alpha^2 \mu_l
^2}{8 \m ^4 \En^2 T^2} \left[ M^4 T- 2 \m^2 T(M^2+4\En T-
4\En^2)-\m M^2 (M^2 +4 \En T- 2T^2)
 \right]\end{equation}
where $\mu_l$ is the neutrino magnetic moment in units of $\mu_B$ (see 
equation (4)).
It can easily be checked that in the limit of $M \rightarrow 0$, we get
the equation for a massless right handed neutrino as found in
\cite{vogel}.

To obtain the differential cross section for LED models, one
needs to take into account the tower of KK final states allowed
by kinematic considerations. Furthermore to compare experiments,
we need to take into into account the flux of incoming neutrinos
with different energies. For this purpose, we choose the
$^{235}U$ reactor flux in the parameterization given in
\cite{vogel}. Folding all these effects in, we get
\begin{equation} \left\langle \frac{d \sigma}{d T} \right\rangle
=\sum_{\mathrm{all M}} \int_ {\En^{\mathrm{min}} (T, M)} ^\infty
\frac{d N_{\nu}}{d \En} \frac{d \sigma(\En, \mathrm{M, T})}{d T}
d\En\end{equation} where $\frac{d N_{\nu}}{d \En}$ is the reactor
spectrum which can be approximated by
\begin{equation}
\frac{d N_{\nu}}{d \En}= \exp(a_0+a_1 \En+ a_2 \En^2).
\end{equation}
We will be considering the case of the $^{235} U$ reactor, which
has the parameters, $a_0=0.870$, $a_1=-0.160$ and $a_2=-0.0910$.

A few comments are in order here. Although we have a spectrum of
all possible incoming neutrino energies, not all of them will
contribute to the scattering at recoil energy, T, and KK mass, M.
This is because energy momentum has to be conserved. For
this reason we do not integrate $\En$ from $0$ to $\infty$.
The lower limit on our integration is
\begin{equation}
\En^{\mathrm{min}} (T, M)=\frac{M^2 + 2\m T}{2(\sqrt{T^2+ 2\m
T}-T)}.
\end{equation}

As for the actual increments in M, we took
\begin{equation}
M=\frac{\sqrt{\sum_{i=1,..,D}n_i ^2 }}{R}, \quad n_i=0,1,2,....
\end{equation}
where $D$ is the number of extra dimensions and the radius of the
extra dimension(s) $R$ is obtained from
\begin{equation}
R=\frac{1}{\pi}\left(\frac{M_{\mathrm{Planck}} ^2}{M_{*D}
^{D+2}}\right)^{1/D}
\end{equation}
and $M_{*D}$ is the $D+2$ dimensional Planck mass and is taken to
be 1 TeV. 

The cross section is sensitive to the mass of the neutrinos since that
determines the value of the magnetic moment $\mu_l$ (see Eq. (4)). In
our calculation presented below, we have chosen
the electron neutrino mass to be $0.6$ eV so as to be compatible with the
oscillation data and the current cosmological limits from WMAP
and SDSS observations. This corresponds to the case where all the
neutrinos are quasi-degenerate in mass. If instead we chose the
normal or the inverted hierarchy, $m_\nu$ would be much smaller
and as a consequence our final results will need to be appropriately
scaled down.

Figures 1,2 and 3 show our results for the case of $D=2$ and $D=3$
as compared to the phenomenological 4D case as well as the purely
weak interaction.

\begin{figure}[h!]
\begin{center}
\epsfxsize15cm\epsffile{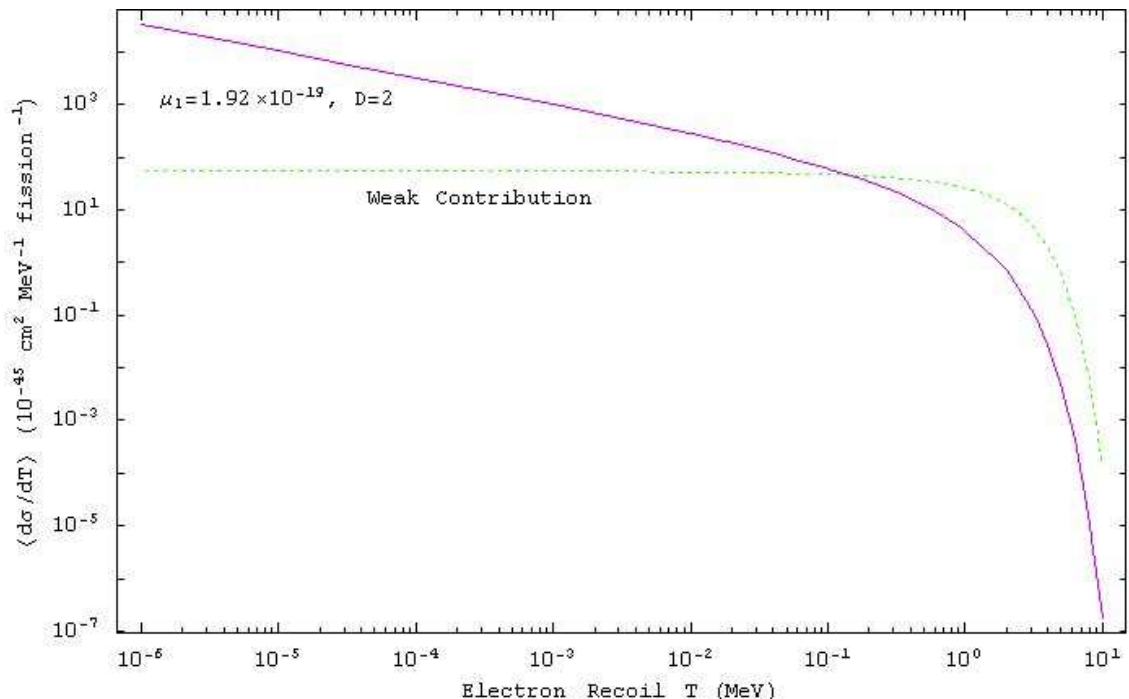}

\caption{The figure shows the folded differential cross section,
$\frac{d\sigma}{dT}$, for the magnetic moment contribution to $\bar{\nu}
e$ scattering for the case of $D=2$ and the standard weak
contribution. The 6D Planck mass is taken to be 1 TeV.
 \label{fig1}} \end{center} \end{figure}

 \begin{figure}[h!]
\begin{center}
\epsfxsize15cm\epsffile{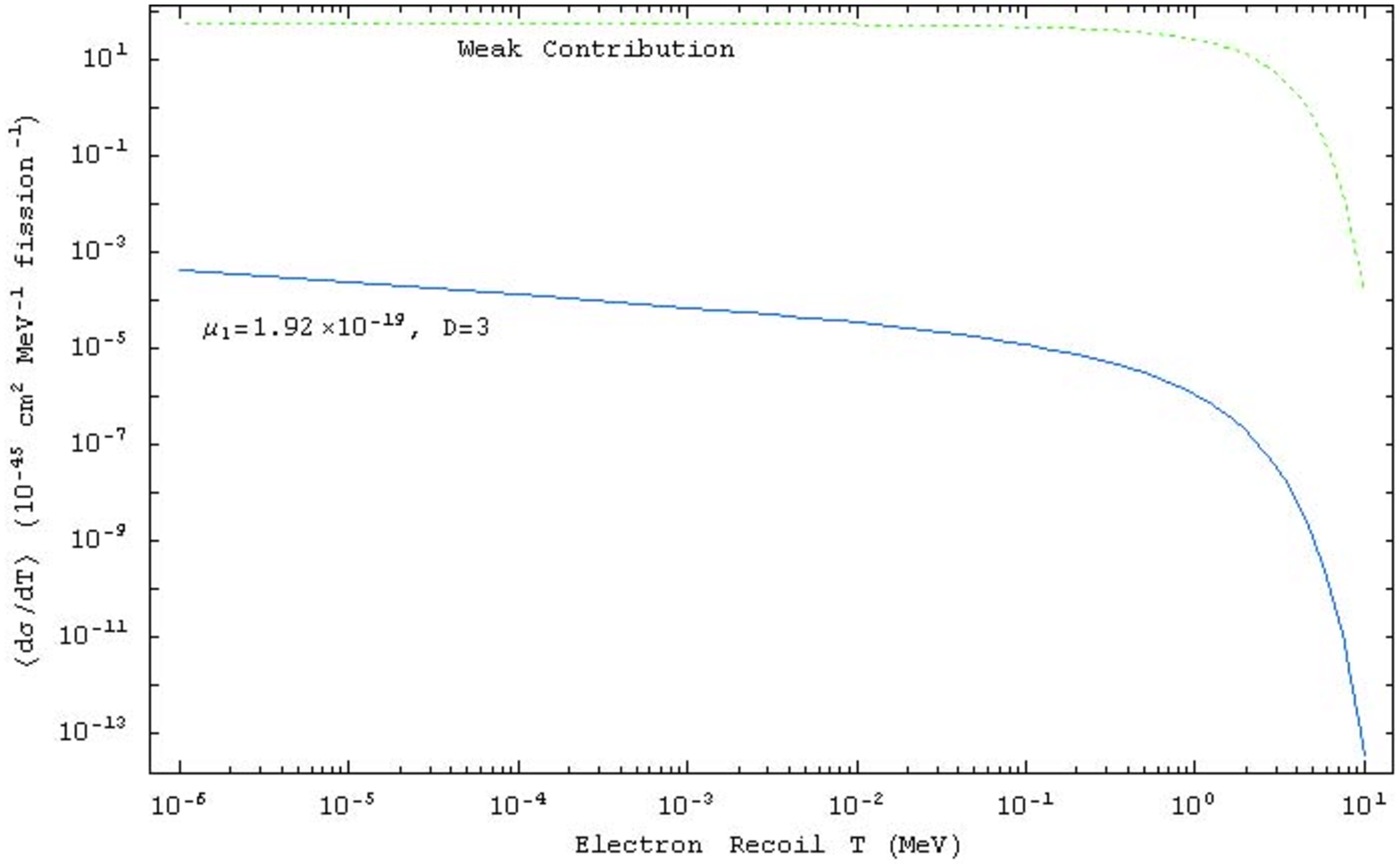}

\caption{The figure shows the folded differential cross section,
$\frac{d\sigma}{dT}$, for the magnetic moment contribution to $\bar{\nu}
e$ scattering for the case of $D=3$ and the standard weak
contribution. The 7D Planck mass is taken to be 1 TeV.
 \label{fig2}} \end{center} \end{figure}

\begin{figure}[h!]
\begin{center}
\epsfxsize15cm\epsffile{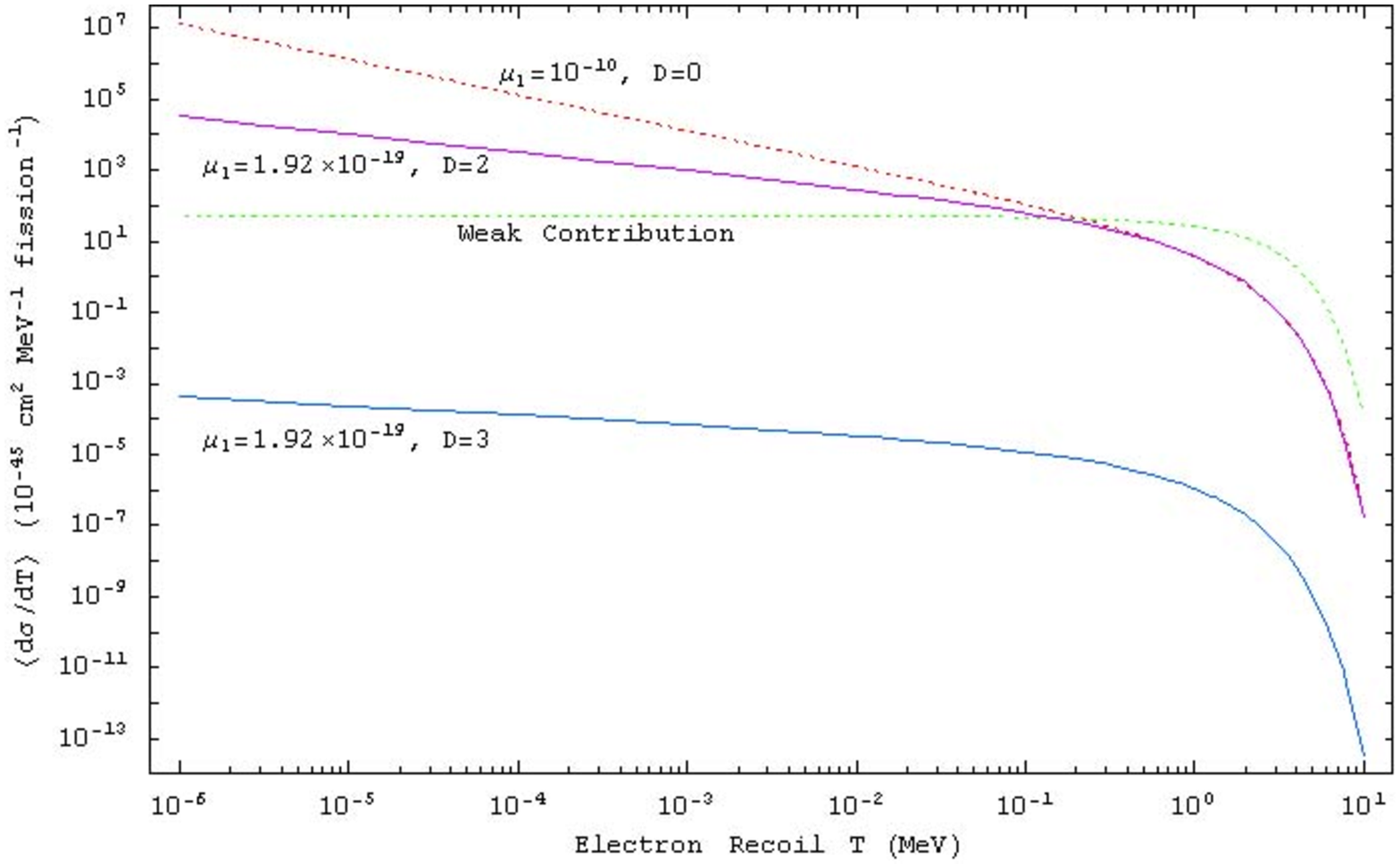}

\caption{The figure presents the folded differential cross section
$\frac{d\sigma}{dT}$ for the magnetic moment contribution to $\bar{\nu}
e$ scattering for the cases of $D=0$, $D=2$, $D=3$ and the
standard weak contribution. The case $D=0$ corresponds to the
standard model extended by the inclusion of on e right handed
neutrino per family.
 \label{fig3}} \end{center} \end{figure}

We see from the above figures that the spectral shape of the
$d\sigma/dT$ for the case of large extra dimensions is different
from the single right hended neutrino case. If the spectral shape
was same as the single RH neutrino case, it would be hard to
search for extra dimensions using the magnetic moment
contribution to $\bar{\nu} e$ scattering. Our results can be used to
put limits on the size of extra dimensions in LED type theories.
This requires a detailed fitting analysis of the theory versus the
observations. It is probably a bit premature to do that since for
our choice of parameters, the predictions are roughly at the same
level as the current experiments. However the proposed reactor
experiments \cite{future} will improve the precision of the data
and can help to constrain the size of the extra dimensions
further. We also point out that even though we have described our
calculation for a model with three bulk neutrinos and Dirac
neutrino masses, it is also valid for LED models which have
Majorana masses for neutrinos\cite{maj}.

In conclusion, we have presented calculations of the magnetic
moment contribution to the $\bar{\nu} e$ scattering cross section for
realistic reactor fluxes for models with large extra dimensions.
We find that for the case of two large extra dimensions, the
predictions of these models is at the level of current
experimental limit on the neutrino magnetic moment. Furthermore,
the spectral shape for the case of LED models is different from
the case of a single right handed neutrino. This can be used to
distinguish the LED case from the pure four dimensional case.
Improvement of the precision of the
reactor measurements of $\bar{\nu}-e$ scattering can therefore provide
useful constraints on the size of  extra hidden space dimensions.

 This work is supported by the National Science Foundation Grant
No. PHY-0099544.

\end{document}